\documentclass[journal=ancac3,manuscript=article]{achemso}
\pdfoutput=1
\usepackage{chemformula} 
\usepackage[T1]{fontenc} 
\usepackage{booktabs}
\usepackage{listings}
\usepackage{lmodern}
\usepackage{mathpazo}
\usepackage{microtype}
\usepackage{graphbox}

\author{Bing Liu}
\altaffiliation{These authors contributed equally to this work}
\affiliation{Physikalisches Institut, Universit\"at W\"urzburg, D-97074 W\"urzburg, Germany}
\alsoaffiliation{W\"urzburg-Dresden Cluster of Excellence ct.qmat, Universit\"at W\"urzburg, D-97074 W\"urzburg, Germany}
\email{bing.liu@physik.uni-wuerzburg.de}

\author{Tim Wagner}
\altaffiliation{These authors contributed equally to this work}
\affiliation{Physikalisches Institut, Universit\"at W\"urzburg, D-97074 W\"urzburg, Germany}
\alsoaffiliation{W\"urzburg-Dresden Cluster of Excellence ct.qmat, Universit\"at W\"urzburg, D-97074 W\"urzburg, Germany}

\author{Stefan Enzner}
\altaffiliation{These authors contributed equally to this work}
\affiliation{Institut f\"ur Theoretische Physik und Astrophysik, Universit\"at W\"urzburg, D-97074 W\"urzburg, Germany}
\alsoaffiliation{W\"urzburg-Dresden Cluster of Excellence ct.qmat, Universit\"at W\"urzburg, D-97074 W\"urzburg, Germany}

\author{Philipp Eck}
\affiliation{Institut f\"ur Theoretische Physik und Astrophysik, Universit\"at W\"urzburg, D-97074 W\"urzburg, Germany}
\alsoaffiliation{W\"urzburg-Dresden Cluster of Excellence ct.qmat, Universit\"at W\"urzburg, D-97074 W\"urzburg, Germany}

\author{Martin Kamp}
\affiliation{Physikalisches Institut, Universit\"at W\"urzburg, D-97074 W\"urzburg, Germany}

\author{Giorgio Sangiovanni}
\email{e-mail: sangiovanni@physik.uni-wuerzburg.de}
\affiliation{Institut f\"ur Theoretische Physik und Astrophysik, Universit\"at W\"urzburg, D-97074 W\"urzburg, Germany}
\alsoaffiliation{W\"urzburg-Dresden Cluster of Excellence ct.qmat, Universit\"at W\"urzburg, D-97074 W\"urzburg, Germany}

\author{Ralph Claessen}
\affiliation{Physikalisches Institut, Universit\"at W\"urzburg, D-97074 W\"urzburg, Germany}
\alsoaffiliation{W\"urzburg-Dresden Cluster of Excellence ct.qmat, Universit\"at W\"urzburg, D-97074 W\"urzburg, Germany}
\email{claessen@physik.uni-wuerzburg.de}

\title[An \textsf{achemso} demo]
  {Moir\'{e} pattern formation in epitaxial growth on a covalent substrate: Sb on InSb(111)A}

\keywords{moir\'{e} pattern, Sb film, InSb(111)A, topological surface state, quasi-particle interference}

\begin{document}

\begin{tocentry}
\includegraphics[width=1.0\linewidth]{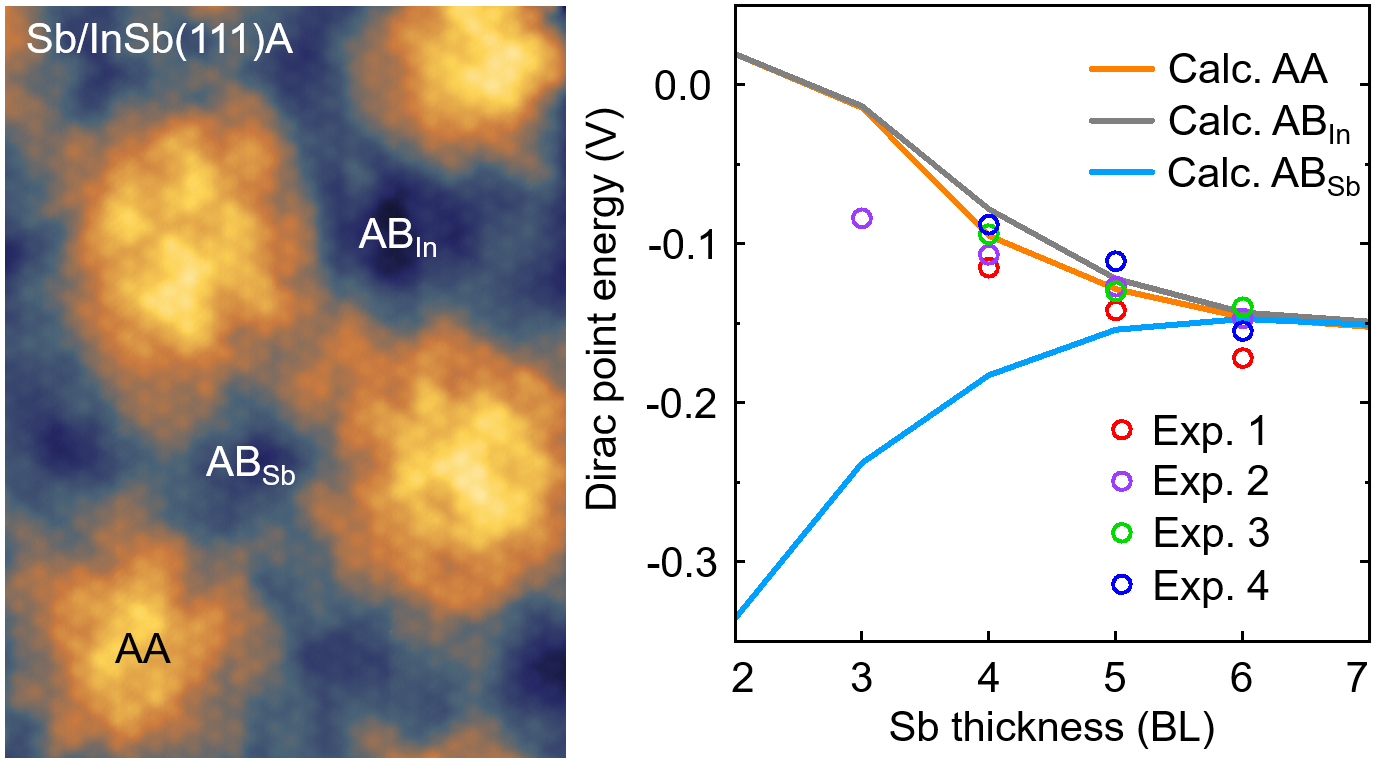}




\end{tocentry}

\begin{abstract}
Structural moir\'{e} superstructures arising from two competing lattices may
lead to unexpected electronic behavior, such as superconductivity or Mottness. 
Most investigated moir\'{e}
heterostructures are based on van der Waals (vdW) materials, as strong interface
interactions typically lead to the formation of strained films or regular
surface reconstructions. Here we successfully synthesize ultrathin Sb films,
that are predicted to show thickness-dependent topological properties, on
semi-insulating InSb(111)A. Despite the covalent nature of the substrate
surface, we prove by scanning transmission electron microscopy (STEM) that already
the first layer of Sb atoms grows completely unstrained, while azimuthally
aligned. Rather than compensating the lattice mismatch of $-6.4\%$ by structural
modifications, the Sb films form a pronounced moir\'{e} pattern as we evidence by
scanning tunneling microscopy (STM) topography up to film thicknesses of several
bilayers. Our model calculations based on density functional theory (DFT) assign the
moir\'{e} pattern to a periodic surface corrugation. In agreement with DFT
predictions, irrespective of the moir\'{e} modulation, the topological surface
state known on thick Sb film is experimentally confirmed to persist down to low film thicknesses,
and the Dirac point shifts towards lower binding energies with decreasing Sb
thickness.
\end{abstract}

\section{Introduction}

Recently the exotic properties in moir\'{e} superstructure have attracted
immense research attention in the field of material physics. These can be
constructed via twisting angles or lattice mismatching at the interface of
layered materials.\cite{cao2020tunable,pierce2021unconventional,liu2021moire} Up
to now, the vast majority of studied moir\'{e} systems are based on
two-dimensional (2D) materials with interlayer van der Waals (vdW)
interactions,\cite{geim2013van} such as twisted bilayer graphene
(TBG),\cite{cao2018unconventional,nimbalkar2020opportunities} graphene/BN
heterostructures,\cite{dean2013hofstadter} and transition-metal dichalcogenide
(TMDC)
heterostructures.\cite{Zhang2017a,tang2020simulation,seyler2019signatures}
Especially the band structure of TBG can be tuned by rotational twisting,
inducing superconductivity or Mott insulating
behavior.\cite{cao2018unconventional,yankowitz2019tuning,po2018origin,chen2021realization}
Contrary to vdW materials, covalent materials are terminated with unsaturated
dangling bonds on the surface,\cite{gatos1961dangling} usually resulting in
strong interface interaction with epitaxial films on top.\cite{xu2018gapped}
Therefore, it is challenging to prepare moir\'{e} superstructures in a 2D film
on a covalent substrate,\cite{bian2022dative} however it may result in
extraordinary properties, especially for a topological film.

Antimony (Sb), a material featuring strong spin-orbit coupling, has been widely
investigated and proven to be a three-dimensional (3D) topological semimetal by
experiment and theory.\cite{Hsieh2009, Seo2010, Zhang2012a} Despite coexisting
with bulk carriers, its spin-polarized two-dimensional (2D) surface states are
topologically protected, which has been supported by the experimental
observation of high transmission rate through surface barriers such as step
edges.\cite{Seo2010} Decreasing the thickness of a 3D topological insulator (TI)
is known to induce a gap opening in the 2D topological surface states through
the hybridization between the opposite surfaces (i.e., between surface and
substrate interface) below a critical thickness.\cite{Li2010,Zhang2010} Indeed,
first principle calculations for free-standing Sb films predict a series of
phase transitions upon reducing their thickness towards the few-layer regime:
from the 3D topological semi-metal for bulk Sb to a 3D TI, further to a quantum
spin Hall (QSH) phase, and finally to a topologically trivial semiconductor in
the case of $\beta$-antimonene, a single buckled honeycomb bilayer (BL) of
Sb.\cite{Zhang2012a}

The interaction with the inevitable substrate will affect the topological states
of antimony films in real systems. So far, Sb films have been realized on
a plethora of substrates: metal surfaces (e.g.,
Ag(111),\cite{Mao2018,Shao2018,Sun2020} Cu(111)\cite{Zhu2019}), 3D TIs (e.g.,
Bi$_2$Se$_3$,\cite{Jin2016,Holtgrewe2020,Hogan2019}
Bi$_2$Te$_3$,\cite{Jin2016,Lei2016} Sb$_2$Te$_3$,\cite{Jin2016,Lei2016}
Bi$_2$Te$_2$Se\cite{Jin2016,Kim2016}), or semiconductor surfaces (Bi-modified
Si(111),\cite{Yao2013} Ge(111)\cite{fortin2017synthesis}). Most of these
substrates strain the epitaxial antimony film to the corresponding substrate
lattice and strongly tune its electronic properties by, e.g., interfacial charge
transfer or even a topological proximity effect.\cite{Kim2016,Holtgrewe2020} In
contrast, Sb films grown on the covalent Ge(111) surface are unstrained, thereby
forming an interfacial moir\'{e} pattern.\cite{fortin2017synthesis} However, the
mechanism underlying this remarkable observation has not been studied in any
detail so far. An additional interesting question concerns the effect of the
moir\'{e} structure on the topological surface states of the Sb film.

Here, we report on the first successful preparation of Sb films on an InSb(111)A
substrate by molecular beam epitaxy (MBE), with film thicknesses ranging from
the few-layer limit to more than $> 20$ BL. Despite the covalent nature of the
InSb(111) surface and very similar to the case of Ge(111), already the first Sb
BL grows unstrained, i.e., does not adapt to the substrate lattice but rather
assumes the lattice constant of bulk Sb, as revealed by scanning transmission
electron microscopy (STEM). As seen by scanning tunneling microscopy (STM), the
corresponding lattice mismatch at the interface induces a moir\'{e} pattern with
an in-plane unit cell size of $\approx6.85\ $nm, which matches integer multiples
of the lattice constants of substrate and films ($15$ and $16$, respectively).

In order to address the large moir\'{e} supercell with first principles
calculations, we have locally approximated the stacking configuration for distinct
high-symmetry sites of the supercell. In agreement with experimental evidence,
these density functional theory (DFT) calculations show that the moir\'{e}
pattern can be attributed to a surface corrugation of the Sb films which is
observable even for film thicknesses of several BLs. The calculations demonstrate that the unstrained configuration is energetically more stable than a model in which the Sb film is strained to the InSb(111) substrate. Additionally, DFT reveals that a
topological surface state, which has been proven on bulk Sb,\cite{Hsieh2009, Seo2010} is preserved above a
critical thickness. This is experimentally confirmed by the observation of
quasi-particle interference probed by scanning tunneling spectroscopy (STS).
Experiment and theory also agree on a shift of the surface state's Dirac point towards lower binding energies with decreasing film thickness. 

\section{Results and discussion}



\begin{figure}[hbt]
\includegraphics[width=0.56\linewidth]{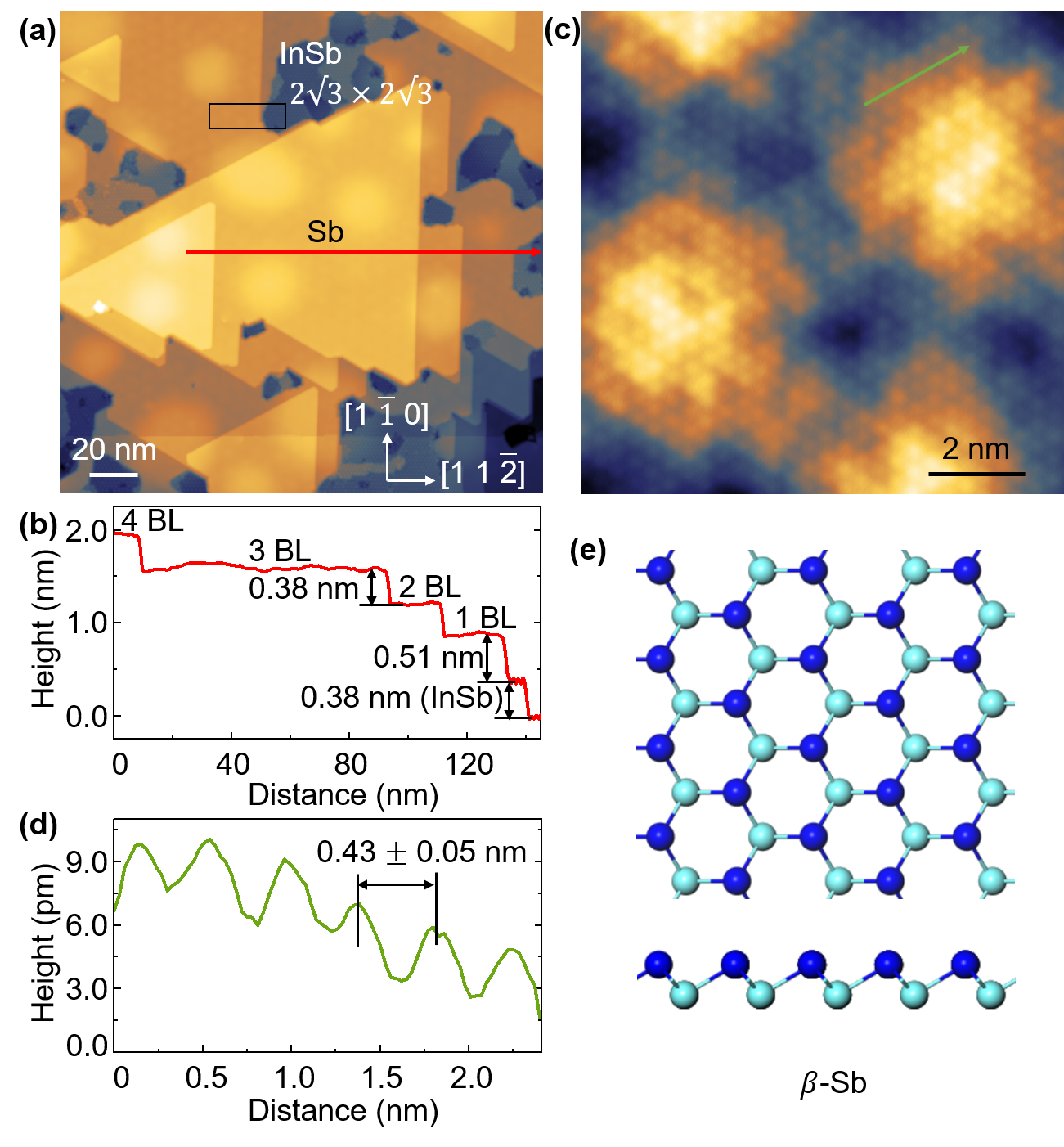}
\caption{
Topography of few-layer Sb on InSb(111)A. (a) Large-scale STM image ($1.5\ $V,
$30\ $pA) of few-layer antimony on InSb(111)A substrate, showing triangular Sb
islands partially covering the $2\sqrt{3}\times2\sqrt{3}$ reconstructed
InSb(111)A surface. (b) Line-profile alone the red arrow in (a). (c)
High-resolution STM image ($100\ $mV, $200\ $pA) of a $5$ BL Sb island, showing
the atomic lattice and superimposed moir\'{e} pattern. (d) Line-profile alone
the green arrow in (c). (e) Top and side views of the atomic model of
antimonene. The dark and light blue atoms represent the upper and lower Sb atoms
in the bucked honeycomb structure, respectively.
}
\label{fig: figure1}
\end{figure}

For the epitaxial growth of our Sb films we start
from a clean InSb(111)A (i.e., indium-terminated) substrate surface of $2\times2$ reconstruction (see Figure S1), obtained by cycles of Ar$^{+}$ sputtering and a subsequent anneal to $340^{\circ}$C. After Sb deposition, islands with thicknesses ranging from $1$ BL to $>20$ BL,
dependent on the duration of evaporation, form on the InSb(111)A surface (Figure
S2). While the outline of the first Sb BL is mostly arbitrary, all further
layers crystallize in a pyramid-like triangular shape. Around the Sb islands, a
Sb wetting layer forms which can be removed without affecting the Sb islands by
post-annealing to $200^{\circ}$C (Figure \ref{fig: figure1}a). The anneal
exposes the clean InSb(111)A substrate surface in a Sb-rich surface
reconstruction of $2\sqrt{3}\times2\sqrt{3}$ periodicity.\cite{Nishizawa1998}
The line profile across several steps of the Sb layers in Figure \ref{fig:
figure1}b allows to determine the step height between two Sb terraces to $0.38\
$nm, in good agreement to the single atomic step height of bulk Sb in the [111]
orientation~\cite{Chekmazov2015}. The step height of the first Sb layer to the
substrate is $0.51\ $nm. This difference could be attributed to either a
different bond length between substrate and Sb layers compared to the bond
length between Sb layers or to the fact that STM does not show the actual
topography of the surface but rather a mixture of topography and the local
density of states (LDOS). The right most step with height of 0.38 nm comes from
the substrate, which can be identified from the obvious
$2\sqrt{3}\times2\sqrt{3}$ reconstruction on top and also corresponds well to
the known single-step height height of InSb(111).\cite{li2020epitaxial} Figure
\ref{fig: figure1}c shows the close-up image of a $5$ BL terrace with atomic
resolution of the Sb lattice and an additional superimposed intensity modulation
that can be attributed to the moir\'{e} pattern. The symmetry and lattice
constant of the surface unit cell of $(0.43 \pm 0.05)\ $nm (Figure \ref{fig:
figure1}d) is identical to that of the (111) surface of a pristine Sb
crystal.\cite{Chekmazov2015} Therefore, the Sb films are determined to be
composed of vertically stacked $\beta$-antimonene layers (called antimonene in
the remainder of this paper), in which the Sb atoms form a buckled honeycomb
lattice as depicted in Figure \ref{fig: figure1}e.


\begin{figure*}[hbt]
\includegraphics[width=1.0\linewidth]{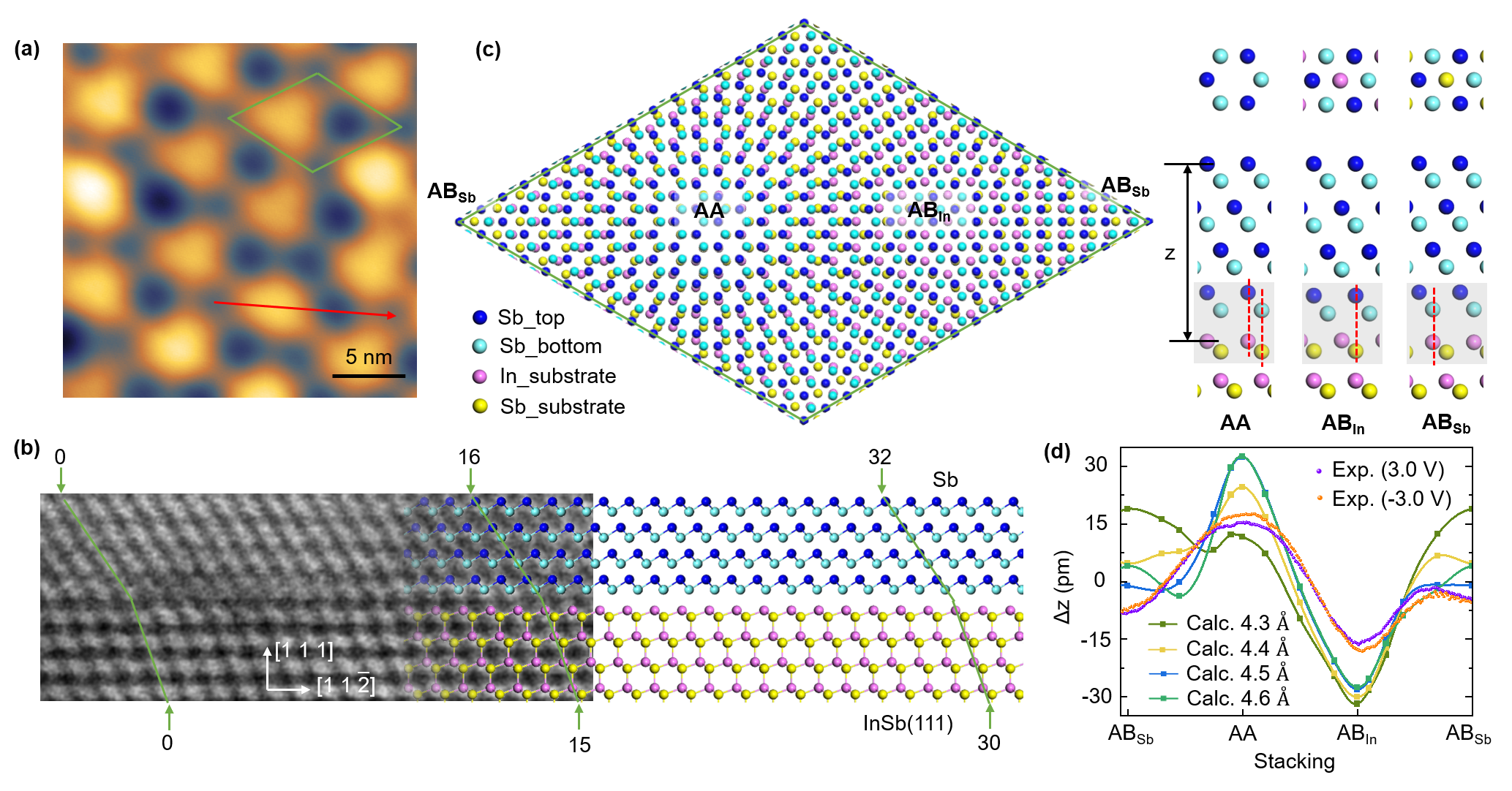}
\caption{
Moir\'{e} pattern in the few-layer Sb/InSb(111)A system. (a) STM image ($3.0\
$V, $100\ $pA) of $4$ BL Sb on InSb(111)A, showing a regular hexagonal moir\'{e}
pattern with a period of $6.85\ $nm. (b) Cross-sectional STEM image (left panel)
in $[1 \overline{1} 0]$ direction, partially overlaid with the atomic model (right panel). The lattice constants of Sb film and InSb(111)A substrate are acquired by STM, and then orthogonally projected onto the $[1 1 \overline{2}]$ direction by a factor of $\frac{\sqrt{3}}{2}$.
Evidently, 16 unit cells of the Sb layers match with $15$ unit cells of InSb(111)A and
create a moir\'{e} situation at the interface. (c) Left panel: schematic model of the
moir\'{e} pattern of antimonene on the top layer of the substrate. The atomic interface geometries of high local symmetry are marked by AA, AB$_{\text{In}}$, and
AB$_{\text{Sb}}$, respectively. Right panel: magnified top and side views of the three local stacking configurations of
the high-symmetry sites. Here the top views correspond to $1$ BL Sb on the top layer of the substrate, i.e. the shadow areas in the side views. (d) Experimental and calculated (for various lattice constants) atom height profile $\Delta z$ across a moir\'{e} unit cell from AB$_{\text{Sb}}$ to the next AB$_{\text{Sb}}$ position on a 4 BL island (red arrow in (a)). $\Delta z$ is the 
local modulation of the total height $z$ (marked in (c)) with respect to its average value $z_{ave}$ in the moir\'{e} unit cell, i.e., $\Delta z = z-z_{ave}$.
}
\label{fig: figure2}
\end{figure*}

For further insight into the origin of the moir\'{e} superstructure, Figure
\ref{fig: figure2}a shows an STM image of the hexagonal moir\'{e} lattice, which
is acquired on a $4$ BL Sb island. Each moir\'{e} unit cell (green rhombus)
shows one section of maximal intensity and two sections with local intensity
minima, similar to the moir\'{e} superstructures in TBG and TMDC
systems.\cite{Xie2019,Bistritzer2011,Zhang2017a} Cross-sectional STEM is used to
obtain the side view of the interface, shown in Figure \ref{fig: figure2}b. It
is apparent that the Sb film does not adapt to the periodicity of the substrate
but rather assumes the lattice constant of its bulk structure. This finding is
rather unusual for a covalent substrate, like InSb(111)A, which possesses
energetically unfavorable unsaturated dangling bond states at its In-terminated
surface that are prone to form covalent bonds. The strong tendency of Sb to bond
with itself rather than with a covalent substrate has already been reported for
Sb films on Ge(111) which also show a lattice mismatch induced moir\'{e}
pattern.\cite{fortin2017synthesis} Such behaviour can be understood by
considering the energy required to strain free-standing Sb layers to the larger
(smaller) lattice constant of the InSb(111) (Ge(111)) substrates. Our 1x1
calculations indicate that adopting to the InSb(111) lattice constant with the
ideal stacking position is energetically less favorable for the Sb films than
retaining the free-standing lattice constant (see section III in Supporting Information). Interestingly,
bismuth (Bi), another group V element, was also recently reported to grow on
InSb(111) without adapting to the substrates lattice.\cite{Liu2021} However,
rather than forming a moir\'{e} structure as in the case of Sb, the ultra thin
Bi films were found to relieve their lattice mismatch-induced strain by forming
a Sierpi\'{n}ski structure. These two different ways of strain relief suggest
that compared to Bi, Sb forms stronger intra-layer bonds, which keep the bulk
structure of Sb fully intact already from the first bilayer.


It is noteworthy, that after Sb deposition the substrate shows a $1\times1$ bulk
lattice configuration at the interface, as seen by STEM, which indicates that
the Sb growth suppresses the $2\times2$ surface reconstruction of the clean
substrate and stabilizes the atomic configuration
of its bulk. Therefore, even though there is no adaptation of the lattice
structures, an interaction of both materials at the interface takes place. STEM
unambiguously shows that the lattice mismatch-induced moir\'{e} pattern forms
immediately at the film-substrate interface, i.e., already at the first Sb
bilayer, rather than between two adjacent Sb layers within the film.
Additionally, the lattices of the substrate and Sb films are azimuthally
aligned to each other, i.e., the moir\'{e} pattern is not generated by a
rotational misalignment of substrate and Sb films. In total, $16$ Sb lattice
constants ($6.87\ $nm) cover $15$ lattice constants of InSb(111)A ($6.88\ $nm),
forming the moir\'{e} superlattice of $\sim 6.85\ $nm seen in STM. Further
confirmation for the rotational alignment of the lattices is provided by the STM
topographies which simultaneously show the atomic lattices of the Sb films and
of the substrate (Figure \ref{fig: figure1}a and Figure S4).

Figure \ref{fig: figure2}c shows the atomic representation of the moir\'{e}
situation at the interface in top view which is limited, for the sake of
clarity, to a single layer of antimonene on top of the InSb(111)A surface layer.
In this moir\'{e} supercell, three local atomic geometries of distinct
symmetries can be identified, labeled here as AA, AB$_{\text{In}}$ and
AB$_{\text{Sb}}$. The notation is adapted from that used for bilayer graphene
and TMDCs \cite{Zhang2017a, Kim2017} and denotes by the index which atom species
of the lower lattice remains uncovered by the upper lattice in top view. For AA,
the atomic positions of both lattices match while for AB$_{\text{In}}$
(AB$_{\text{Sb}}$), the In (Sb) substrate atoms are (in top view) centered in
the honeycombs of the first antimonene bilayer of the film (see schematic atomic
structure in Figure 2c).

For a better understanding of the interfacial bonding situation and the
mechanism underlying the moir\'{e} pattern formation we now turn to a
theoretical analysis. We will start by looking at the different local bonding
situations, i.e., by performing DFT calculations for simple $1\times1$ lattices
whose unit cell represents the various interface alignments. This approach explicitly excludes interactions between the different stacking configurations in the full moir\'{e} lattice. 
The analysis of the
latter would require the simulation of the full Sb on InSb moir\'{e} unit cell,
for which a brute-force systematic \textit{ab-initio} simulation is computationally out
of reach.
Despite the local approximation, these calculations allow for a surprisingly accurate inspection of the stacking-dependent electronic and structural properties, as we demonstrate in the following.

By comparison of the relaxed atomic structures of different sites inside the
moir\'{e} unit cell, it becomes evident that the height of the Sb
adatoms above the substrate (visualized in Figure \ref{fig: figure2}d) varies
for different stacking configurations. In our primitive cell calculations,
substrate and film possess the same in-plane lattice parameter for which we
assumed various values between those of bulk InSb(111) ($4.6\ $\AA) and bulk Sb ($4.3\ $\AA). In Figure 2d
we show the resulting theoretical height profiles for a $4$ BL Sb film in
comparison to their experimental counterparts. These profiles are taken along
the horizontal diagonal of a moir\'{e} unit cell (cf. Figure 2c), covering all
high-symmetry sites (i.e., from AB$_{\text{Sb}}$ on the left corner via AA and
AB$_{\text{In}}$ to the next AB$_{\text{Sb}}$ on the right corner). 
Both experiment and theory find a maximal (minimal) corrugation at the AA (AB$_{\text{In}}$) stacking. The height for the AB$_{\text{Sb}}$ alignment is instead sensitively affected by the lattice constant, which is plausible as this is the portion of the unit cell experiencing the strongest interface interaction (more detail further below). In contrast to the experimental moir\'{e} structure, the bonding described by this simplified approach strongly depends on the film's and substrate’s local strain. Notwithstanding this limitation, it is able to satisfactorily describe the overall trend in the observed structural pattern. 

In this context we wish to point out that the STM data is robust against a variation of
the tunneling voltage (purple and orange curves in Figure 2d are acquired at
$3.0\ $V and $-3.0\ $V, respectively), demonstrating that the measured line
profile is dominated by the surface topography and not by variations in the
local density of states. A similar height trend of the calculated interface Sb
layer, i.e., the first BL Sb film, is shown in Figure S5. These results show
that the moir\'{e} pattern manifests itself as an atomic height modulation
between interface and first Sb layer, which translates to surface corrugations even for
several bilayers, similar to the Sb/Ge(111) case.\cite{fortin2017synthesis} As
expected, the corrugation of the moir\'{e} pattern decreases with increasing
layer thickness (see Figures S2, S8, and S9). The overall good agreement between theory
and experiment suggests that the moir\'{e} surface corrugation is mostly driven
by the local stacking energies, with long-range interactions being of only minor
importance.


	\begin{table}
		\begin{center}
	   	\caption{
DFT calculations of the simplified local 1$\times$1 geometries for the bond length at the interface between the first Sb layer and the top layer of the
substrate, as well as the Sb-Sb intralayer bond with lattice constant of 4.3 \AA. Sb$_{\text{bottom}}$ (Sb$_{\text{top}}$) is the bottom (top) Sb atom of the first
Sb layer within their buckled honeycomb structure, and In (Sb) is the In (Sb)
atom of the top InSb(111)A layer.
}
			\begin{tabular}{c | c | c | c}
				bond length [nm]		& AA 		&AB$_{\text{In}}$	& AB$_{\text{Sb}}$	\\
				\hline
				Sb$_{\text{bottom}}-$In		& $0.40$	& $0.35$			& $0.30$			\\
				Sb$_{\text{bottom}}-$Sb		& $0.38$	& $0.42$			& $0.46$			\\
				Sb$_{\text{top}}-$In			& $0.44$	& $0.47$			& $0.52$			\\
				Sb$_{\text{top}}-$Sb			& $0.59$	& $0.50$			& $0.60$			\\
				\hline
				Sb$_{\text{bottom}}-$Sb$_{\text{top}}$& $0.30$	& $0.30$			& $0.29$			\\

			\end{tabular}
			\label{tab: table1}
		\end{center}
	\end{table}


\begin{figure}[hbt]
\includegraphics[align=c,width=0.56\linewidth]{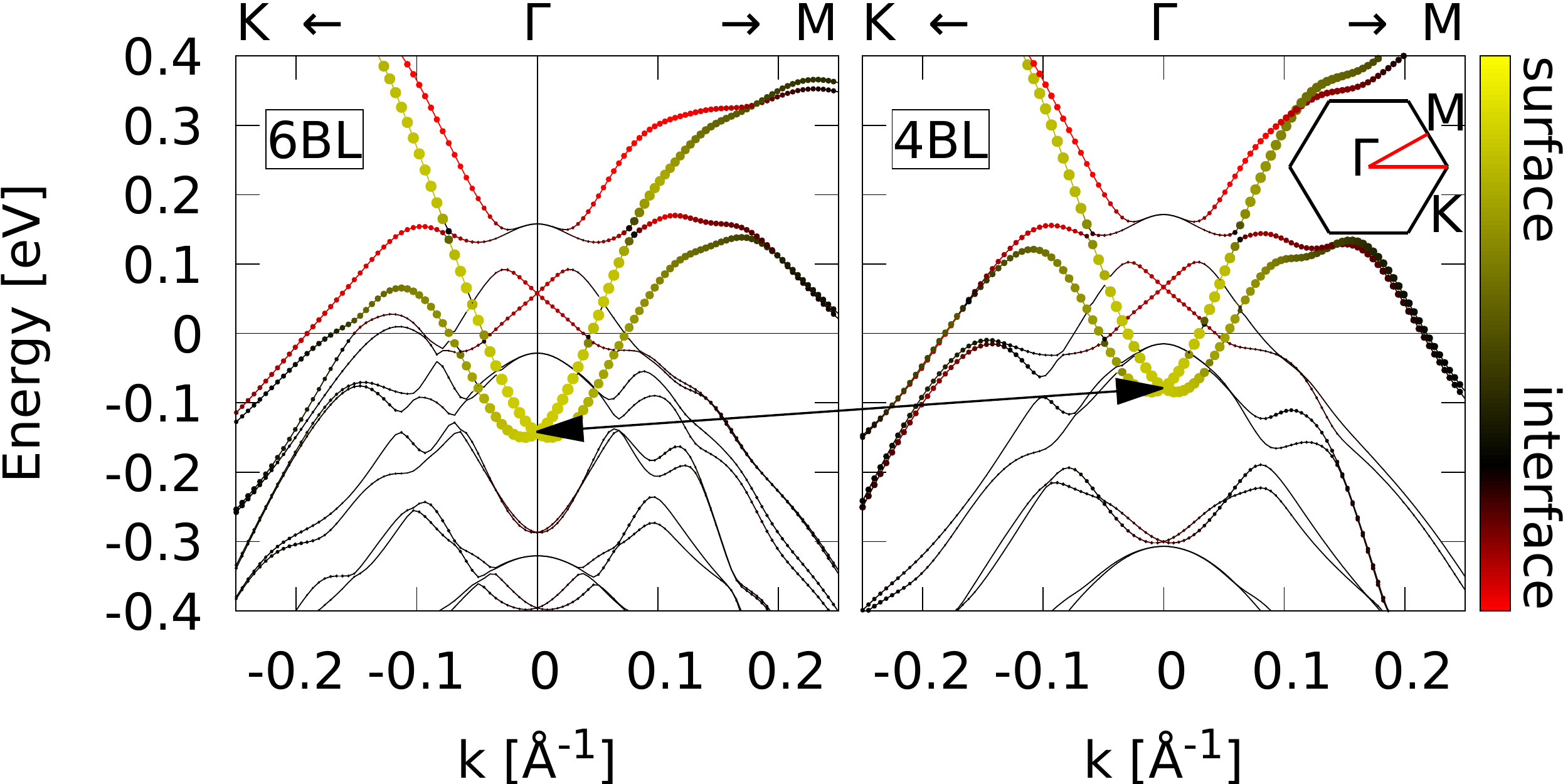}
\caption{Band structures along the K$-\Gamma-$M paths obtained by DFT for Sb layers on InSb(111)A substrate in AB$_{\text{In}}$ stacking. The point size is given by combined projection weight of both surface and interface Sb BL and thus allows to distinguish these states from InSb and bulkier Sb bands. The top surface state indicated by the yellow color shifts to higher energies with respect to the Fermi level as the thickness of the film decreases from 6 BL to 4 BL, whereas the red colored interface bands remain at the same energy.
}
\label{fig: figure3}
\end{figure}

The local atomic models for the $4$ BL Sb films on InSb(111) additionally allow
us to take a closer look at the bond length between the substrate and the Sb
films at the high-symmetry sites (Table \ref{tab: table1}). For AA and
AB$_{\text{In}}$, the shortest bond between substrate and Sb film is found to be
$0.38\ $nm (Sb atom of substrate to the bottom Sb atom of Sb layer) and $0.35\
$nm (In atom of substrate to the bottom Sb atom of Sb layer), respectively. Both
are much larger than the intra-bilayer Sb-Sb bonds ($0.30\ $nm), indicating a
weak interaction at the interface. In contrast, the AB$_{\text{Sb}}$ geometry
corresponds to a regular continuation of InSb(111) with the bottom Sb atom on
top of the In atom at the interface, resulting in the shortest interfacial bond
length ($0.30\ $nm) within the moir\'{e} supercell. For stackings close to
AB$_{\text{Sb}}$, the dominating interface bond is close to the Sb intra-bilayer
distances of 0.29 nm, hereby indicating a much stronger interaction with the
substrate.
	
Having established the local substrate-Sb bonding conditions, the influence on
the electronic properties will be investigated in the following with the focus
on the Sb-surfaces. The surface states of Sb thick films connect conduction and valence states. With decreasing thickness, the surface and interface film layers interact more strongly, resulting in a hybridization gap and Rashba-type surface states (see Figures S6 and S7).
This transition occurs below thicknesses of 4 and 5 BL depending on the high-symmetry
position, determined by the interplay of substrate/interface-Sb interaction and Sb surface/interface
interaction (see Figures S6 and S7).

As shown in the band structure of Figure \ref{fig: figure3} for
AB$_{\text{In}}$, the top Sb surface state (indicated in yellow) moves down w.r.t.
to the Fermi level with increasing film thickness. The same height-dependent trend
is also found for most of the sites inside the moir\'{e} supercell including the
high-symmetry AA layering (Figures S6 and S7). For stackings close to
AB$_{\text{Sb}}$ the opposite trend is observed, where the Sb surface state
shifts up in energy with increasing thickness (see Figures S6 and S7). In
contrast, the energy positions of the interface states colored with red in
Figures 3, S6 and S7 are determined by the stacking configuration and are barely
influenced by the film thickness.


\begin{figure}[htbp]
\includegraphics[width=0.56\linewidth]{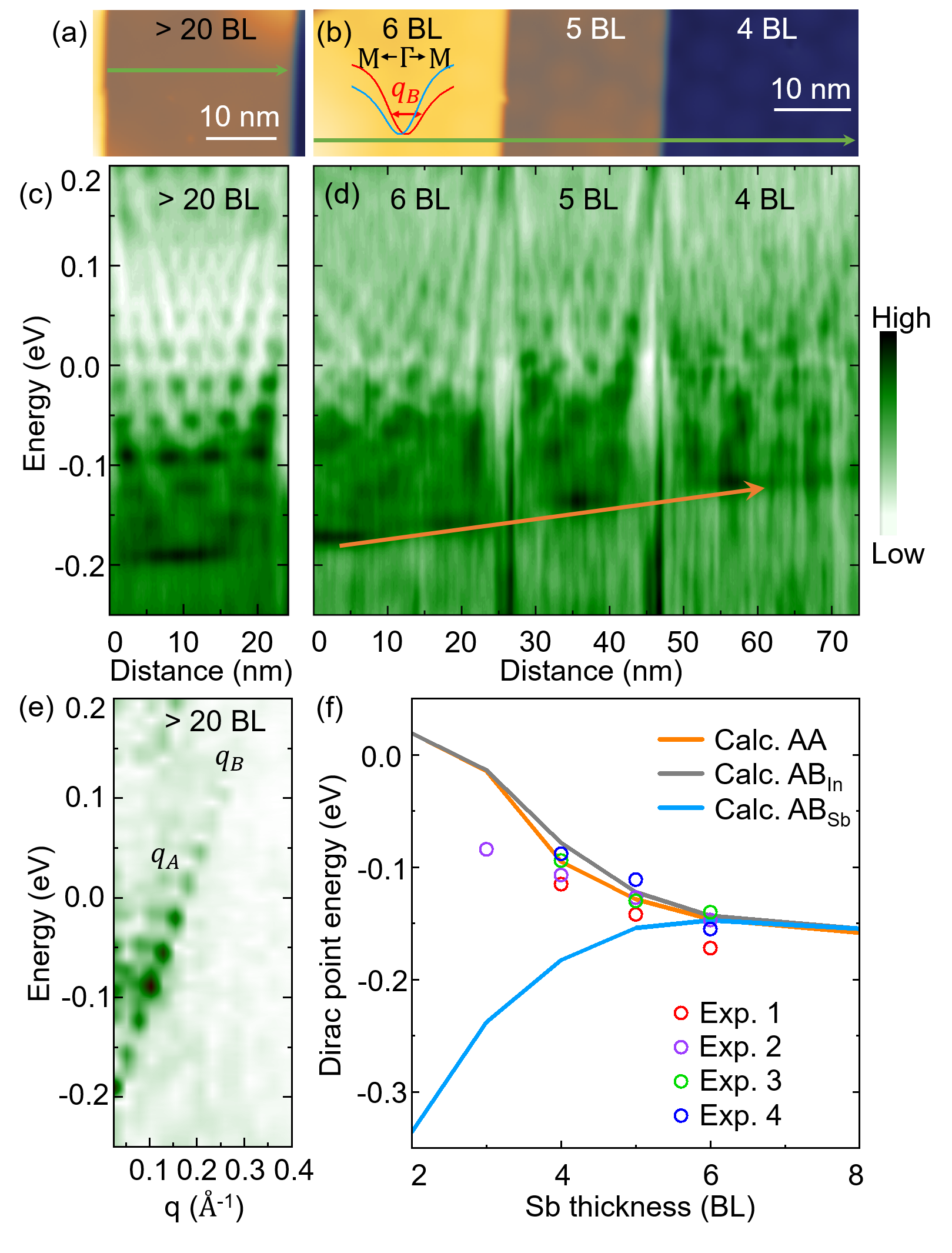}
\caption{Thickness-dependent STS study of Sb layers on InSb(111). (a) STM image ($1.0\
$V, $10\ $pA) of a terrace of a thick Sb layer on InSb(111)A. (b) STM image ($1.0\ $V, $10\ $pA)
of terraces of $4$ BL to $6$ BL Sb on InSb(111)A. Insert: the schematic of the
Sb surface state at $\Gamma$, and the scattering wave vector \emph{q}$_{\text{B}}$ (red
arrow). (c) Spatially resolved spectra along the green arrow in (a), showing interference pattern 
due to surface state scattering at the step edges. (d) STS spectra taken along the green arrow 
in (b). (e) Fourier transform of the line spectra across the thick layer ($\textgreater 20$ BL) Sb shown in (c), 
resulting in quantization of the scattering wavevectors \emph{q}$_{\text{A}}$ and \emph{q}$_{\text{B}}$. 
(f) The summarized energy of the Dirac point experimentally measured in various areas and calculated with different stackings.
}
\label{fig: figure4}
\end{figure}
	
In order to study the thickness-dependent evolution of the electronic structure
of Sb films experimentally, we performed scanning tunneling spectroscopy (STS)
measurements. Spectra are acquired on Sb terraces of varying thickness ($>20$ BL,
$6-4$ BL) along lines perpendicular to Sb step edges, as depicted by green
arrows in Figures \ref{fig: figure4}a and b. The corresponding spatially
resolved variation of the local density of state (LDOS) is shown in Figures
\ref{fig: figure4}c and d, respectively. The spectra taken on a thick Sb film ($
>20$ BL) show a standing wave pattern similar to that observed on the (111)
surface of a single crystal of Sb by Seo \emph{et al}.\cite{Seo2010} For bulk
Sb, this quasiparticle interference (QPI) pattern is generated by the scattering
of the two-dimensional topological surface state from adjacent parallel step
edges. The dominant scattering vector \emph{q}$_{\text{B}}$ is shown in the
inset in Figure \ref{fig: figure4}b (red arrow) and connects surface states with
the same spin (see the detailed scattering geometry in Ref. \citenum{Seo2010}). A fast Fourier transform (FFT) (Figure \ref{fig: figure4}e) of
the spectra in Figures \ref{fig: figure4}c confirms that the scattering features
coincide with that observed for the topological surface state on bulk Sb. For lower film thickness, the interference pattern is still visible but become blurrier and the background intensity stronger (Figure \ref{fig:
figure4}d). We attribute this trend to the increasing surface corrugation of the
moir\'{e} pattern towards lower thickness, which may induce additional
scattering centers for the surface state. The fact that the QPI is still visible
for few-layer Sb films indicates that a surface state, similar to that of bulk
Sb, is preserved despite the presence of a substrate.

Finally, the QPI pattern also allows to extract the position of the surface
state's Dirac point. For the thick Sb film ($>20$ BL), the Dirac point is
located at $-193\ $meV, slightly higher than that of bulk Sb $\approx -225\
$meV, which may be attributed to the influence of the substrate. For $6$ BL, $5$
BL, and $4$ BL Sb, we find the Dirac point at $-172\ $mV, $-143\ $mV, and $-115\
$mV, respectively, indicating a shift towards lower binding energies with
decreasing thickness (see Figure \ref{fig: figure4}f for the experimental data
measured in various sample areas). This thickness-dependent trend of the Dirac
point is also predicted by our DFT calculations for most local stacking
configurations with topological nontrivial surface states, i.e., AA,
AB$_{\text{In}}$, and other non-high-symmetry stackings, while calculations
close to AB$_{\text{Sb}}$ stackings show the opposite trend. However, as stated
before, AB$_{\text{Sb}}$ is a unique local area that shows rather strong
substrate-Sb interaction and particular electronic properties inside the
moir\'{e} unit cell. According to the experimentally observed trend of the Dirac
point and the calculation results, we can assume, that the Sb films of $\geq 4$
BL on InSb(111)A preserve the general topological properties theoretically
predicted for most sites of the film, typified by AA and AB$_{\text{In}}$ sites.

Unfortunately, due to the rather 3D growth which prevents thickness-dependent
angle-resolved photoelectron spectroscopy (ARPES) studies and the influence of
the moir\'{e} pattern on Sb films of low thickness, our experimental data can not
provide information about a thickness-dependent topological phase transition as
predicted by DFT calculations. On the other hand, moir\'{e} superstructures are
recently extensively studied in twisted bilayer graphene or TMDCs due to their
rich physics, such as flat bands and quantum-confined electronic
states\cite{Cao2018,Cao2018a,Kim2017,Li2021,Pan2018}. A moir\'{e} pattern on a
covalent surface could therefore evoke interesting properties which deserve
further studies with more elaborated theoretical models.

\section{Conclusion}

In summary, we successfully synthesized high-quality Sb films on a covalent
InSb(111)A substrate with thicknesses down to the few-layer limit. The Sb layers
do not adapt their lattice to the covalent substrate, but keep their bulk Sb
lattice structure already from the first deposited bilayer, thereby generating a
lattice-mismatch induced moir\'{e} pattern with a supercell lattice constant of
$6.85\ $nm. Our DFT calculations for specific local stacking geometries occurring
throughout the supercell provide support for an energetic stabilization of the
moir\'{e} pattern compared to the case of strained Sb film. These calculations
also show that the Rashba-type surface states of the few BL films transform
above a critical film thickness into the 2D topological surface states known
from bulk Sb. The existence and thickness evolution of these surface states, i.e., shifting towards lower binding energy as decreasing Sb thickness, are experimentally confirmed by the detection of corresponding QPI patterns using STS. Our work not
only introduces a new substrate for the epitaxy of ultrathin Sb films, but also
provides first insights into the mechanism underlying the surprising occurrence
of a lattice mismatch-induced moir\'{e} structure on a covalent substrate.

\section{Methods}

Sb with high purity (99.9999 $\%$) was evaporated from a Knudsen cell with the substrate held at
$\approx110^{\circ}$C. All STM data was acquired with a commercial Omicron
LT-STM at $4.3\ $K. Cross-sectional lamellae for STEM investigations were
prepared on a Dual-Beam System (FEI Helios Nanolab) equipped with an Omniprobe
micromanipulator. Scanning transmission electron microscopy was performed using
an uncorrected FEI Titan 80-300 operating at $300\ $kV, $100 - 120\ $pA beam
current and a convergence semi-angle of $10\ $mrad. The images were recorded
using a high-angle annular dark-field electron detector.

The theoretical studies of local structures with different stackings and numbers
of Sb layers were carried out with first-principle density functional
calculations as implemented in the Vienna \textit{ab initio} simulation package
(VASP)~\cite{VASP}, within the projector augmented-plane-wave (PAW)
method~\cite{PAW1,PAW2}. For the exchange-correlation potential the PBE-GGA
functional~\cite{PBE} was used, by expanding the Kohn-Sham wave functions into
plane waves up to an energy cutoff of $400\ $eV. We sample the Brillouin zone on
a $21\times21\times1$ Monkhorst-Pack mesh and employ spin-orbit coupling
self-consistently~\cite{SOC_VASP}.

We consider two-dimensional systems of 6 InSb BL with hydrogen-passivated
Sb-termination (B face) and $1$ BL to $8$ BL of Sb on the In-terminated (A face)
surface with different in-plane alignments. To disentangle the electronic states
of both surfaces, a vacuum distance of at least $15\ $\AA$\ $between periodic
replicas in z-direction is applied. The structures are relaxed vertically to the
surface until forces converged below $0.01\ $eV/\AA. The height of the $6$ InSb
bilayers are strained to $18.85\ $\AA$\ $ to exhibit a band gap similar to
calculations of the bulk substrate with hybrid functionals as
HSE06.\cite{HILAL201641} The relative alignment between film and substrate bands
might be influenced by this approach, however the effect of the substrate on the
general trends and behaviour of the surface state should be negligible.

\begin{acknowledgement}

This work has been funded by the Deutsche Forschungsgemeinschaft (DFG, German
Research Foundation) through the W\"urzburg-Dresden Cluster of Excellence
EXC2147 on Complexity and Topology in Quantum Matter \textit{ct.qmat} (Project
ID 390858490) as well as through the Collaborative Research Center SFB 1170
\textit{ToCoTronics} (Project ID 258499086). B.L. gratefully acknowledges
support by the Alexander von Humboldt Foundation, Bonn, Germany. G.S. and S.E.
acknowledge the support from the Deutsche Forschungsgemeinschaft (DFG, German
Science Foundation) through FOR 5249-449872909 (Project P5). S. E., P. E. and G.S. thank Domenico Di Sante for useful discussions.

\end{acknowledgement}

\begin{suppinfo}

The following files are available free of charge.
\begin{itemize}
  \item InSb(111)A surface, as-grown Sb islands, energetic analysis of Sb films on InSb(111)A substrate, orientation of moir\'{e} pattern, calculated bond length, height and band structures of Sb films, moir\'{e} intensity and fast Fourier transform (FFT) results
\end{itemize}

\end{suppinfo}

\bibliography{Sb_InSb_111_A}

\end{document}



\newpage
\subsection{I. InSb(111)A surface}




\begin{figure}[!h]
\includegraphics[width=0.8\linewidth]{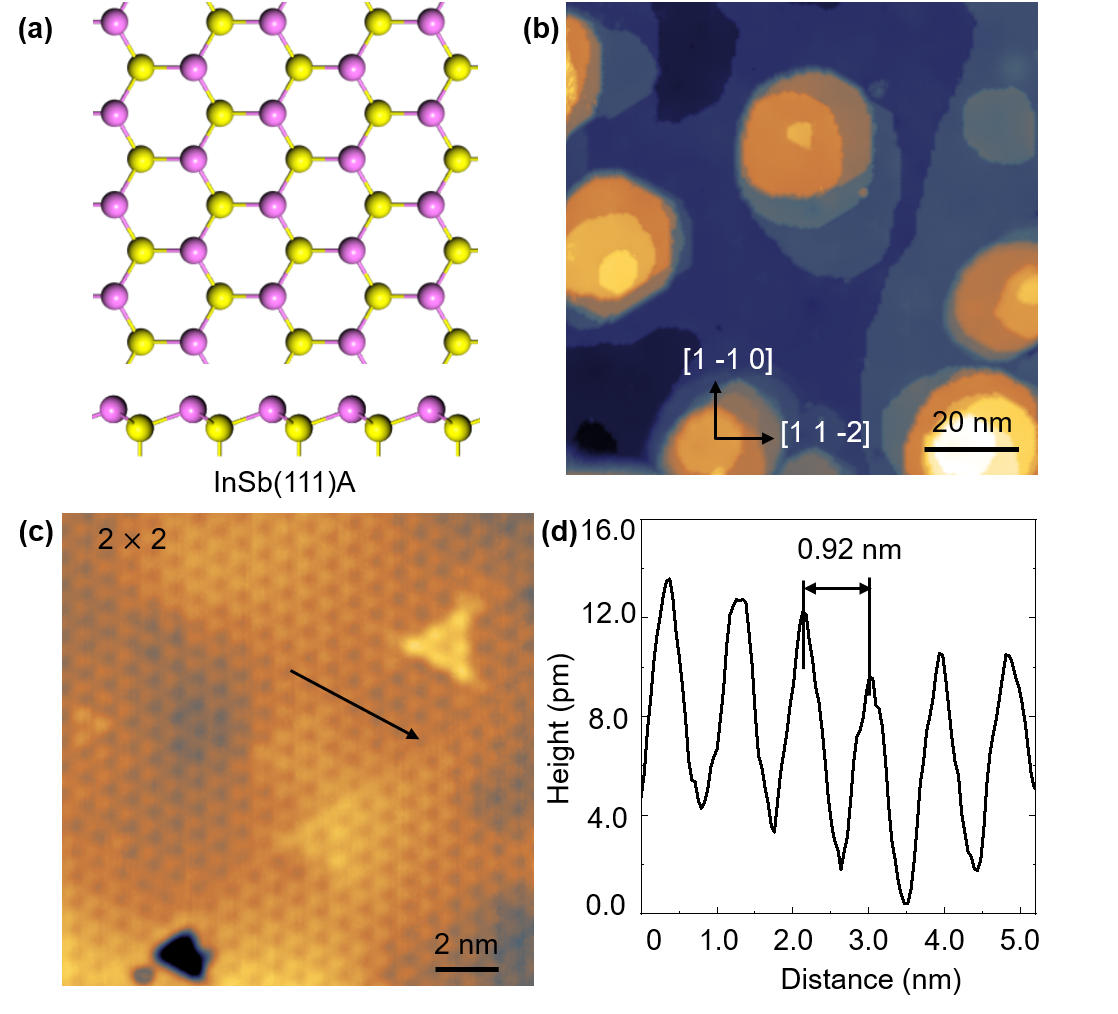}
\caption{
Topography of InSb(111)A surface. (a) Top and side views of the atomic model of InSb(111)A surface. (b) and (c) Large-scale ($1.0\ $V,
$20\ $pA) and atom-resolved ($1.0\ $V,
$50\ $pA) STM images of InSb(111)A surface with $2\times2$ reconstruction, respectively. (d) Line-profile alone the red line in (c), showing the period of a $2\times2$ unite cell is $0.92\ $nm.}
\label{fig: figureS1}
\end{figure}

\newpage
\subsection{II. As-grown Sb islands}


\begin{figure*}[!h]
\includegraphics[width=1.0\linewidth]{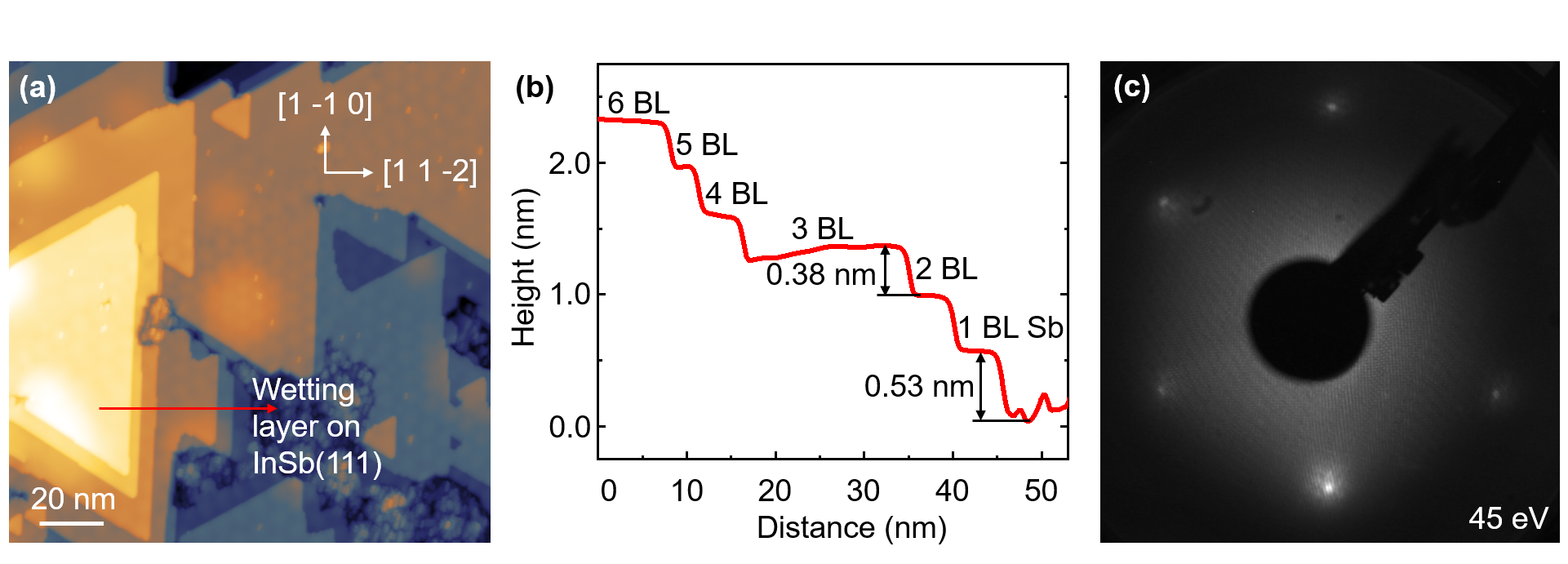}
\caption{
(a) As-grown Sb islands with 1-$7$ BL on InSb(111)A substrate, showing wetting layer around the Sb islands. (b) Line-profile along the red arrow in (a). (c) LEED pattern of a InSb(111)A sample with Sb islands of $1$ to $7$ BL thickness. Moir\'{e}-related satellite spots surrounding the diffraction pattern of Sb(111) surface are visible.
}
\label{fig: figureS2}
\end{figure*}
\clearpage

\subsection{III. Energetic analysis of Sb films on InSb(111)A substrate}	

\begin{figure}[!h]
\includegraphics[width=0.49\linewidth]{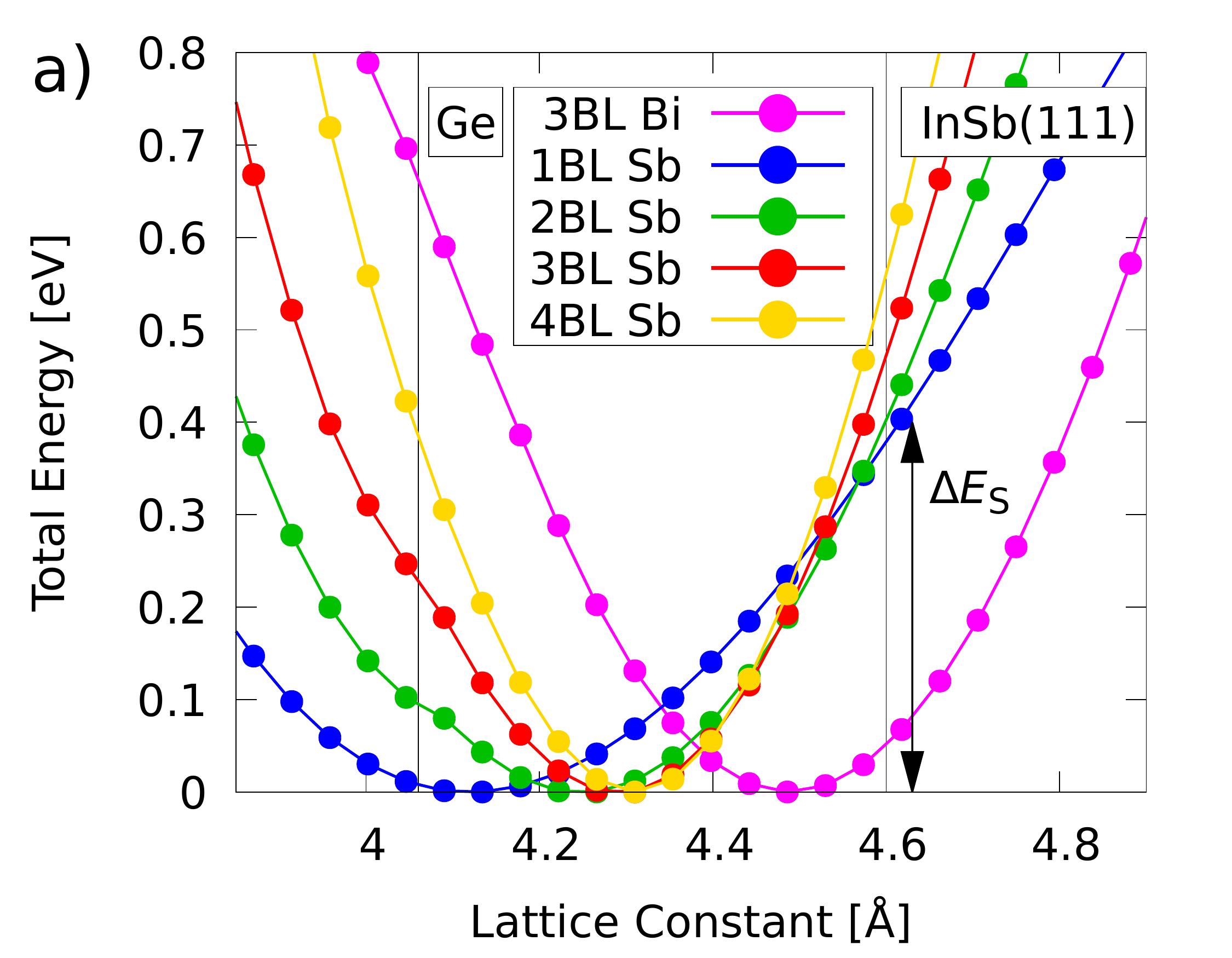}
\includegraphics[width=0.49\linewidth]{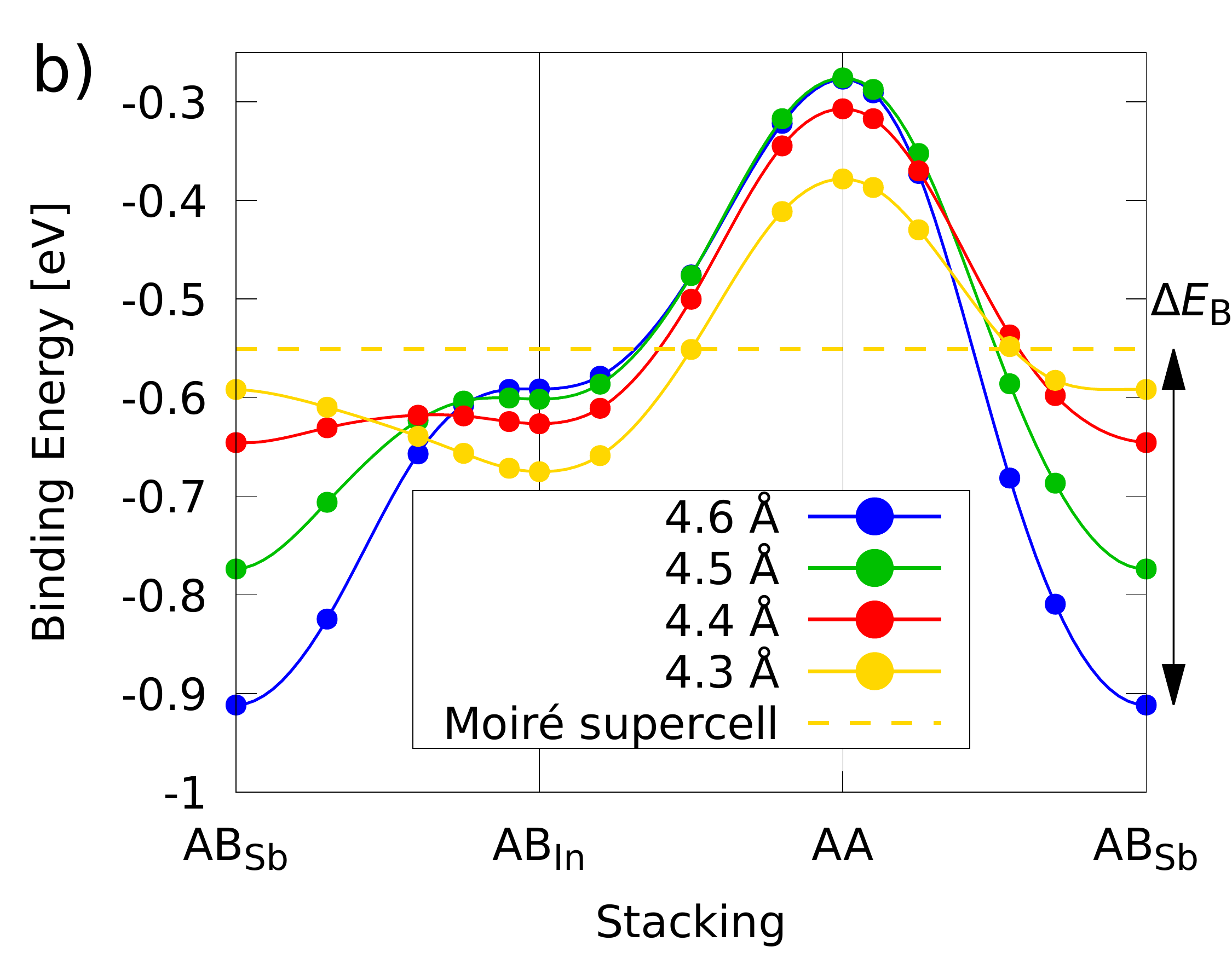}
\caption{
(a) Total energy of strained free-standing Sb bilayers of different thicknesses. Matching the lattice constant of InSb(111) requires more than 0.4\,eV. Additionally the film stiffness increases with thickness.
(b) Binding energies of 4 BL Sb on 6 layers InSb(111) depending on the alignment at the interface and lattice constant. The dashed line indicates an estimate on the binding energy of the moir\'{e} supercell. Even for few Sb bilayers, the unstrained moir\'{e} supercell configuration exhibits lower energies compared to the 1x1 alternative in the most favorable AB$_\text{Sb}$ stacking. The energy $\Delta E_\text{S}$ required to strain exceeds the binding energy gain $\Delta E_\text{B}$ of 4.6\,\AA\ AB$_\text{Sb}$ compared to the 4.3\,\AA\ moir\'{e} structure.}
\label{fig: figureS3}
\end{figure}

To understand the origin of the moir\'{e} structure on InSb(111) we analyzed the total energy of free-standing Sb bilayers and the binding energies for different stackings and lattice constants. The computational details can be obtained from the method section.
The energy $\Delta E_\text{S}$ required to strain the Sb films to the lattice constant of InSb(111) with 4.6 \AA\ is 0.4\,eV for 1 BL Sb and increases with the Sb thickness. As comparison, the total energy of free-standing three bismuth BL is shown in purple, indicating its strain cost with respect to the InSb(111) lattice constant is significantly smaller than for Sb films (see Figure \ref{fig: figureS3}a). \\
The average binding energy of the moir\'{e} structure per primitive cell is estimated by considering the binding energies of various stacking configurations and shown with a yellow dashed line (see Figure \ref{fig: figureS3}b). This is done along the supercell diagonal through the local AB$_\text{In}$, AB$_\text{Sb}$ and AA alignment, similar to the stacking-dependent height analysis. The primitive structures contain 4 BL Sb on 6 BL InSb(111)A. The binding energies are determined for different in-plane Sb lattice constants to compare the commensurate (4.6\AA) and unstrained (4.3\AA) situation. The in-plane lattice constant of the InSb(111) are set to match the Sb film lattice to conduct primitive calculations. For the relaxation the height of the substrate and the in-plane positions are fixed. Film and substrate were separated by more than 15\AA\ and relaxed to determine the binding energy. \\
For the lattice constant of 4.6 \AA, the AB$_\text{Sb}$ stacking exhibits the highest binding energies. The averaged moir\'{e} binding energy $\Delta E_\text{B}$ is around 0.35\,eV lower. However, to reach the strongest bound configuration the Sb film have to be strained to the lattice constant of 4.6\AA. As indicated in Figure \ref{fig: figureS3}a, the in-plane strain requires more energy $\Delta E_\text{S}$ than that can be gained through the binding energy $\Delta E_\text{B}$, even for few Sb layers. The comparison between unstrained averaged supercell and 1x1 commensurate cell hints that it is energetically favorable to minimize the in-plane strain instead of maximizing the binding energy. This indicates strong intra-layer interaction and a weak interaction with the substrate.\\
Even though this analysis does not include moir\'{e} effects beyond the summation of local stackings, the crucial interplay between strain and bonding strength is highlighted.

\newpage
\subsection{IV. Orientation of moir\'{e} pattern}

\begin{figure*}[!h]
\includegraphics[width=1.0\linewidth]{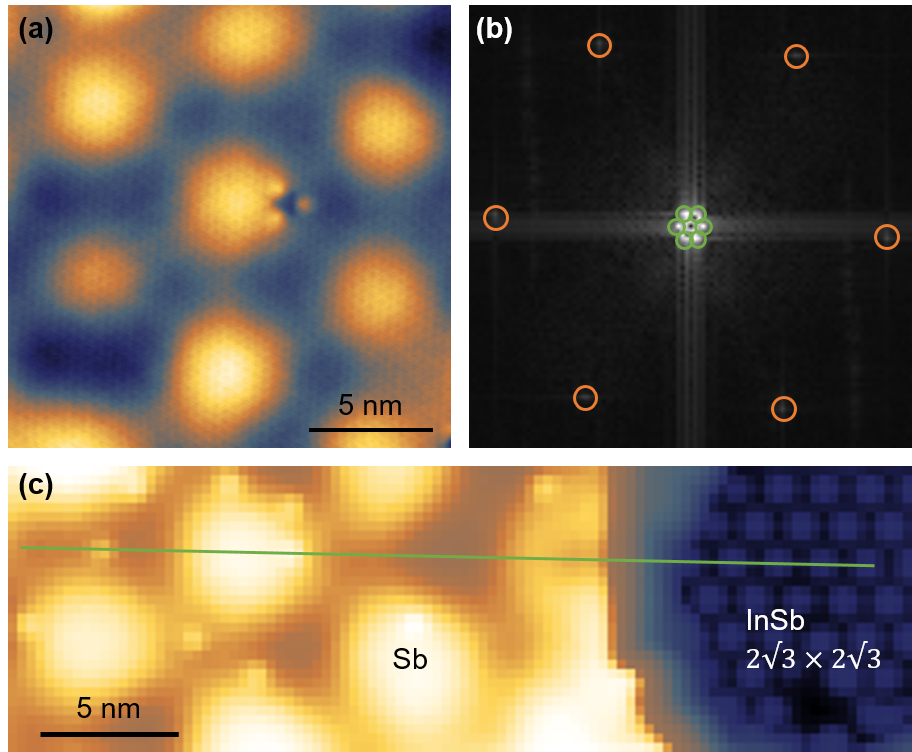}
\caption{
(a) STM image ($0.5\ $V, $500\ $pA) of a $4$ BL Sb on InSb(111)A substrate, showing obviously hexagonal moir\'{e} pattern with a period of $6.85\ $nm. (b) Fast Fourier transform (FFT) patterns of (a). The orange and green circles mark the transformed patterns from Sb and moir\'{e} lattices, respectively. Both STM image and FFT results indicate Sb and moir\'{e} lattices are along the same orientation. (c) Zoom-in image of the black rectangle area in Figure 1a, including both moir\'{e} lattice on Sb films and the $2\sqrt{3}\times2\sqrt{3}$ reconstructed lattice of the substrate. As directed by the green line, moir\'{e} lattice also has the same orientation as the InSb(111)-$1\times1$ structure.
}
\label{fig: figureS4}
\end{figure*}
\clearpage

\subsection{V. Calculated bond length, height, and band structures of Sb films}	

	\begin{table}[!h]
		\begin{center}
			\begin{tabular}{|c | c | c | c | c | c | c | c | c | c|}
			    \hline
			     \multirow{2}{*}{bond length [nm]}   &\multicolumn{3}{|c|}{$4.4\ $\AA$\ $}             &\multicolumn{3}{|c|}{$4.5\ $\AA$\ $}                 &\multicolumn{3}{|c|}{$4.6\ $\AA$\ $}     \\
			    \cline{2-10}
						& AA 	&AB$_{\text{In}}$	& AB$_{\text{Sb}}$   & AA  		&AB$_{\text{In}}$ 	& AB$_{\text{Sb}}$    & AA  	&AB$_{\text{In}}$ 	& AB$_{\text{Sb}}$	\\
				\hline
				Sb$_{bottom}-$In		& $0.39$	& $0.35$			& $0.30$	 	& $0.39$	& $0.35$	& $0.29$        	& $0.40$	& $0.35$			& $0.29$		\\
				\hline
				Sb$_{bottom}-$Sb		& $0.38$	& $0.42$			& $0.46$		& $0.38$	& $0.42$	& $0.46$        	& $0.38$	& $0.42$			& $0.46$		\\
				\hline
				Sb$_{top}-$In			& $0.45$	& $0.47$			& $0.51$		& $0.45$	& $0.47$	& $0.51$        	& $0.44$	& $0.47$			& $0.50$		\\
				\hline
				Sb$_{top}-$Sb			& $0.60$	& $0.49$			& $0.59$		& $0.59$	& $0.48$	& $0.58$        	& $0.59$	& $0.47$			& $0.57$		\\
				\hline
				Sb$_{bottom}-$Sb$_{top}$ & $0.30$	& $0.30$			& $0.29$		& $0.30$		& $0.30$	& $0.29$	   & $0.30$        	& $0.30$	& $0.30$		\\
				\hline

			\end{tabular}
\caption{
Bond length at the interface between the first BL Sb  and the top layer of the
substrate, as well as the Sb-Sb bond of Sb intralayer with lattice constant of $4.4\ $\AA$\ $, $4.5\ $\AA$\ $, and $4.6\ $\AA$\ $, respectively. Sb$_{bottom}$ (Sb$_{top}$) is the bottom (top) Sb atom of the first BL
Sb within their buckled honeycomb structure, and In (Sb) is the In (Sb)
atom of the top InSb(111)A layer.
}
			\label{tab: table1}
		\end{center}
	\end{table}


\begin{figure*}[!h]
\includegraphics[width=1.0\linewidth]{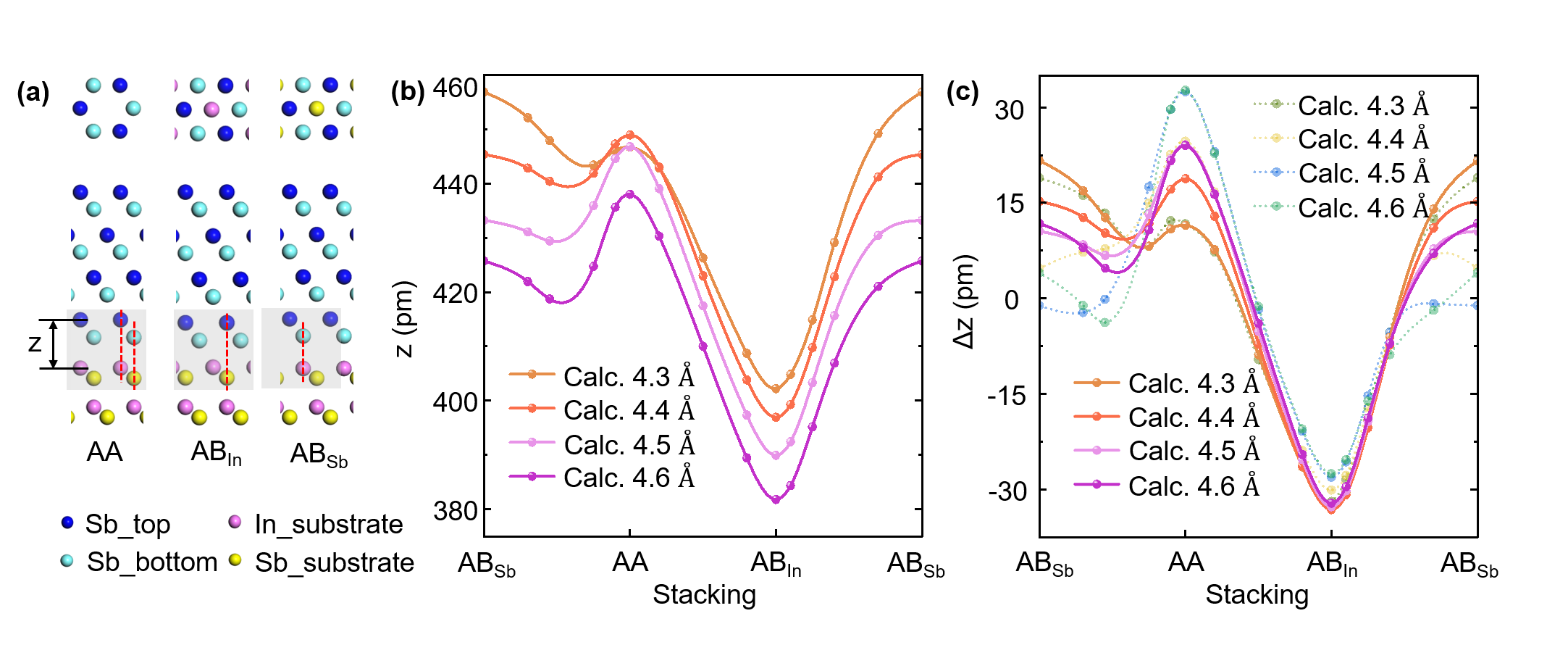}
\caption{
(a) Models for the locally stacking configuration of
the high-symmetry sites of $4$ BL Sb on InSb(111)A. The height of the first BL Sb film to the substrate is marked by $z$. (b) Calculated (with various lattice constants) height-profiles of the first BL Sb film across a moir\'{e} unite cell from AB$_{\text{Sb}}$ to the next AB$_{\text{Sb}}$ position (the black arrow in Figure 2a). The averaged absolute value of $z$ from the first BL Sb film to the substrate is about $4.2\ $\AA, which can compare with the experimentally measured value $5.1\ $\AA\ by STM when the local density of states are also considered. (c) Comparison of the calculated height-profiles ($\Delta z$) of the first BL (solid curves) and $4$ BL (dashed curves) Sb across a moir\'{e} unite cell. For better comparison each line profile is aligned to its average value $z_{ave}$ in the moir\'{e} unit cell, i.e., $\Delta z = z-z_{ave}$. The consistency of the height of both the first BL and $4$ BL Sb indicates the moir\'{e} pattern manifests mainly as a height variation at interface, and translates to corrugations even for several BL Sb.
}
\label{fig: figureS5}
\end{figure*}
		

\begin{figure*}[!h]
\includegraphics[width=1.0\linewidth]{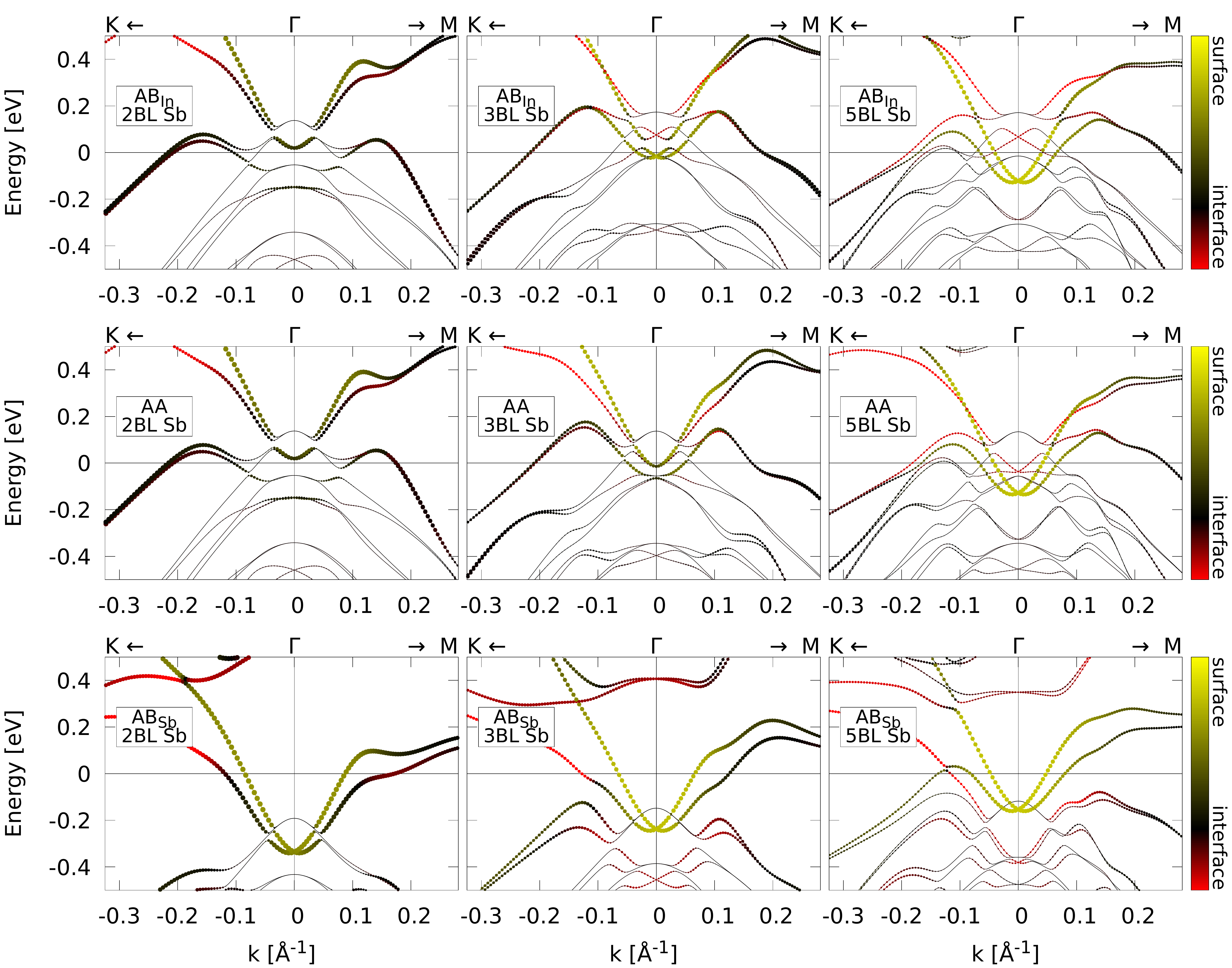}
\caption{
DFT calculated band structures for the $4$ BL, $6$ BL, and $8$ BL Sb on InSb(111)A substrate with AB$_{\text{In}}$, AA, and AB$_{\text{Sb}}$ local configurations, along the K$-\Gamma-$M direction. The bands originating from the orbitals of top (bottom) Sb BL are colored by yellow (red).
}
\label{fig: figureS6}
\end{figure*}

\newpage
\begin{figure*}[!h]
\includegraphics[width=1.0\linewidth]{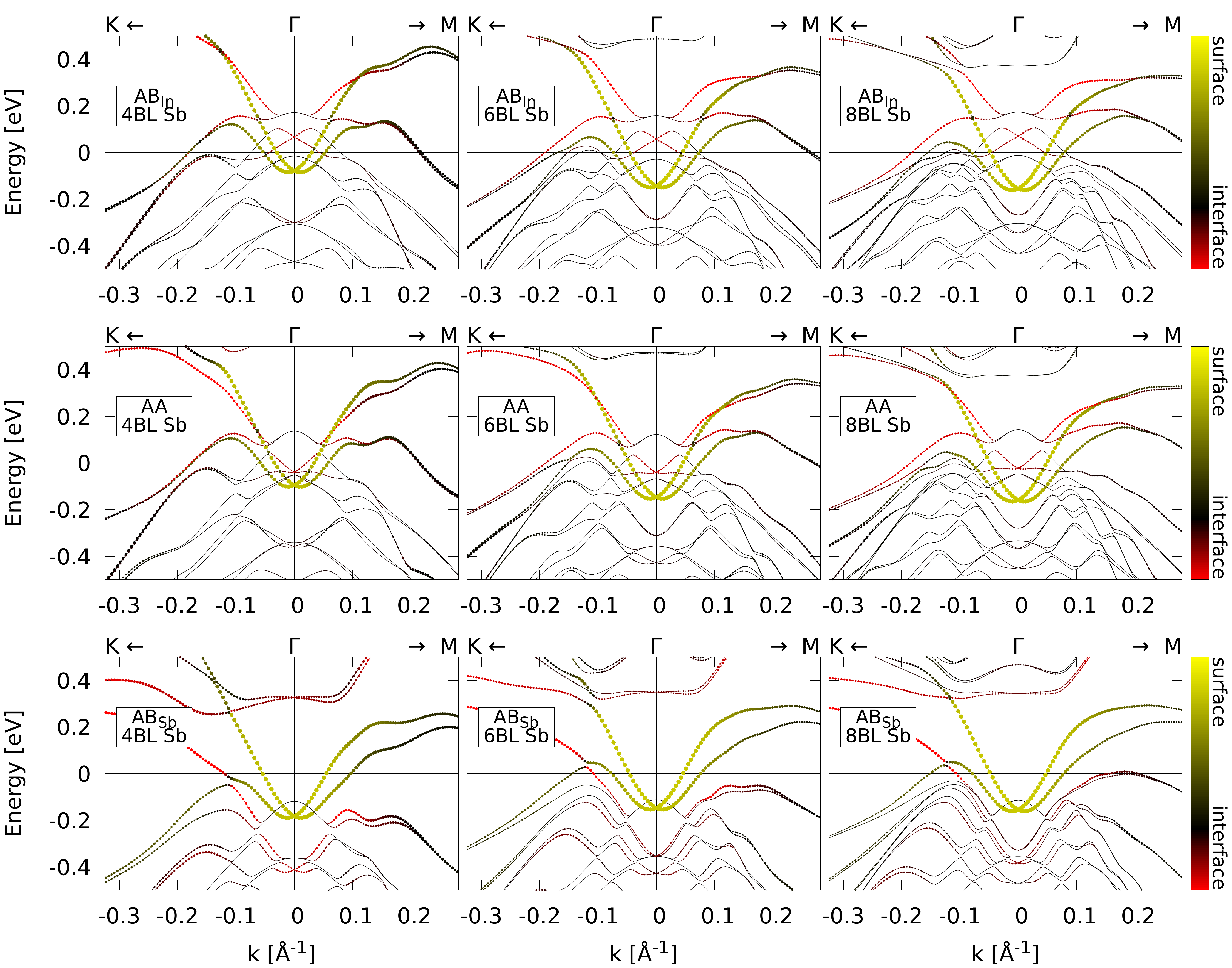}
\caption{
DFT calculated band structures for the $2$ BL, $3$ BL and $5$ BL Sb on InSb(111)A substrate with AB$_{\text{In}}$, AA, and AB$_{\text{Sb}}$ local configurations, along the K$-\Gamma-$M direction. The bands originating from the orbitals of top (bottom) Sb BL are colored by yellow (red).
}
\label{fig: figureS7}
\end{figure*}
\clearpage

\subsection{VI. Moir\'{e} intensity and fast Fourier transform (FFT) results}

\begin{figure*}[!h]
\includegraphics[width=1.0\linewidth]{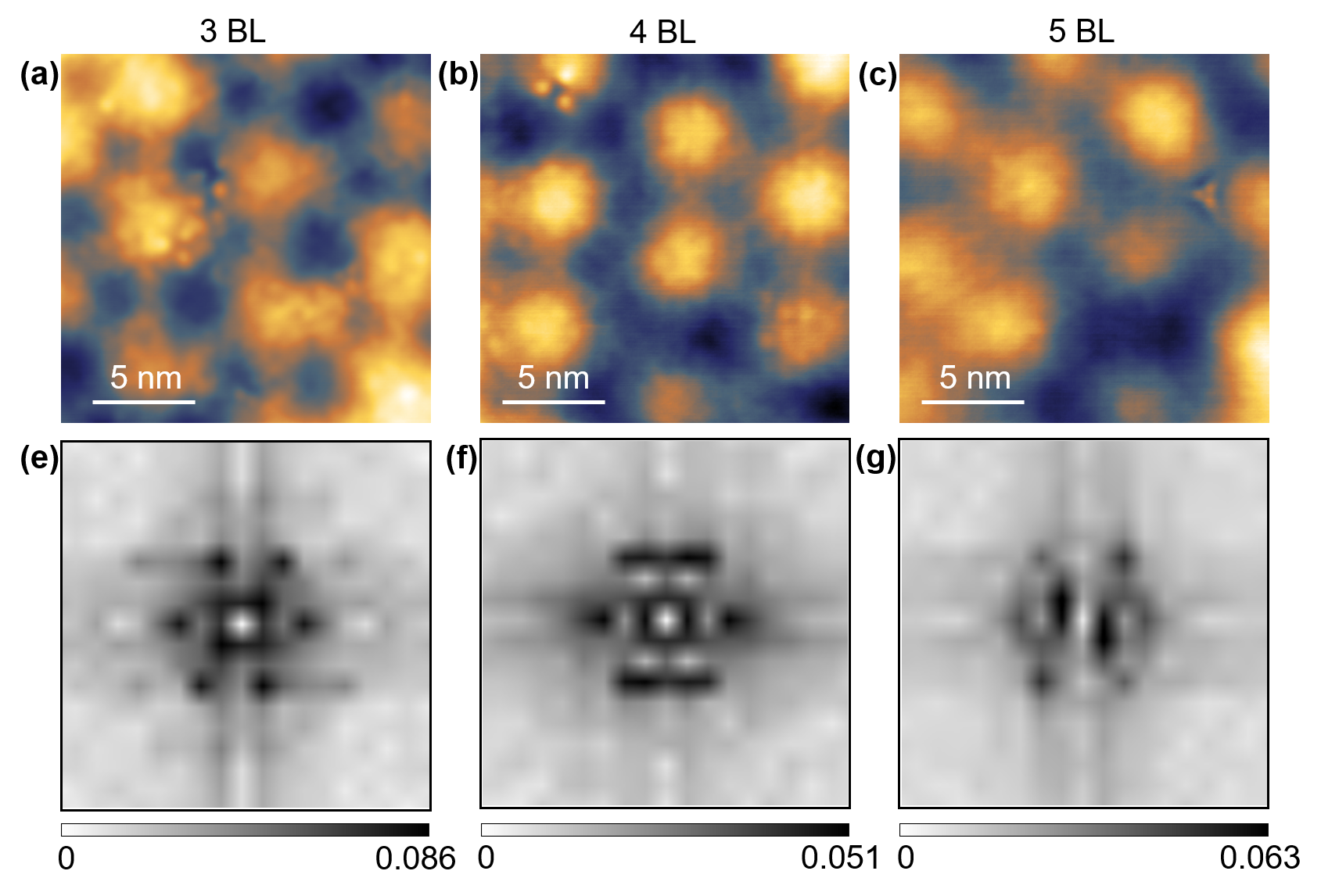}
\caption{
Thickness dependence of moir\'{e} intensity. (a-d) Moir\'{e}-resolved STM images of the $3$-$5$ BL Sb on InSb(111)A substrate. All of the images are acquired at $0.3\ $V,
$200\ $pA. (e-g) Corresponding FFT results of the images in (a-c).
}
\label{fig: figureS8}
\end{figure*}


\begin{figure*}[!h]
\includegraphics[width=1.0\linewidth]{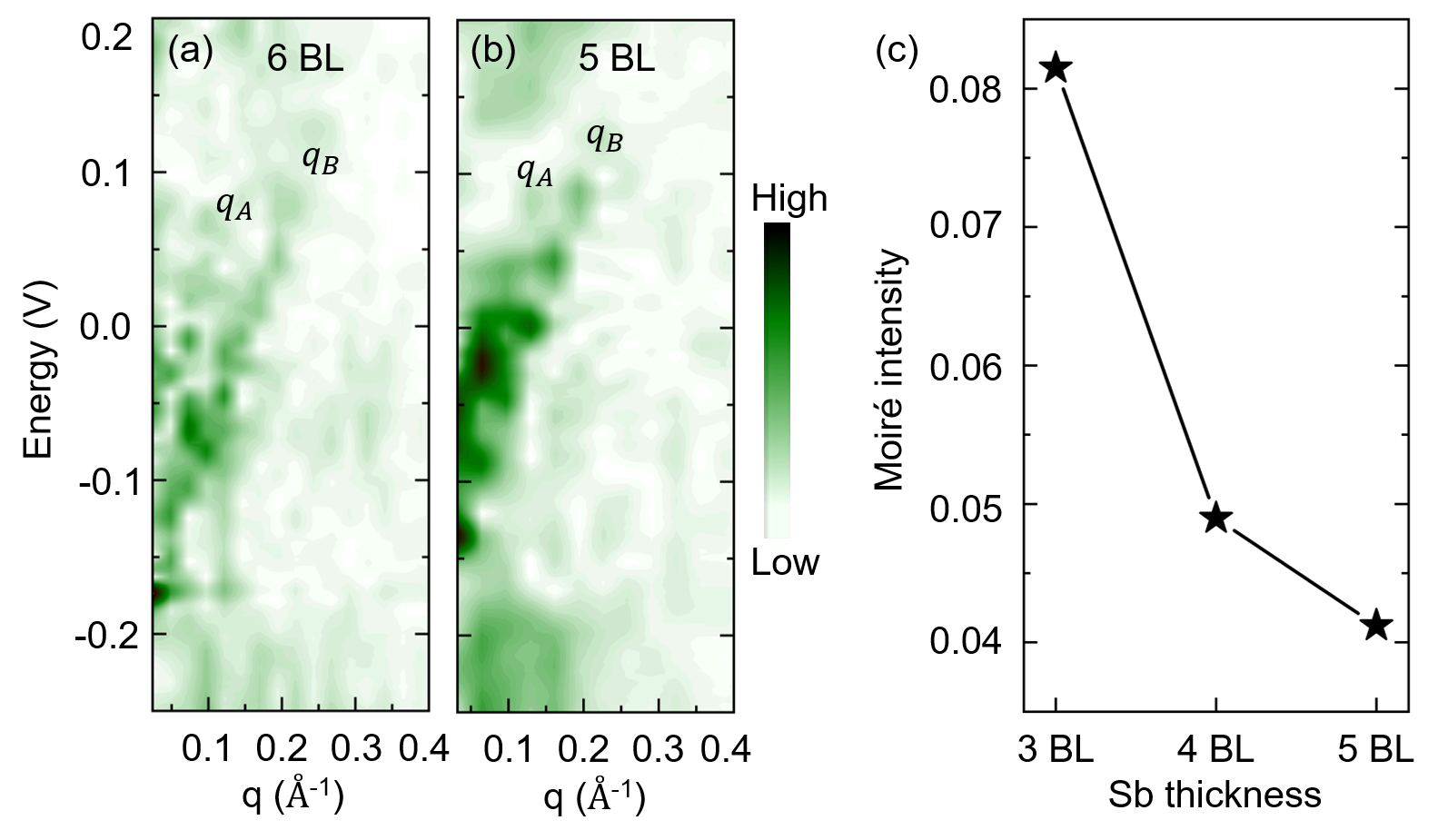}
\caption{
(a and b) Fourier transform of the line spectra across the $5$ BL and $6$ BL Sb shown in Figure 4d, resulting in the quantization of the scattering wavevectors \emph{q}$_{\text{A}}$ and \emph{q}$_{\text{B}}$. (d) Averaged intensity of the moir\'{e}-related six FFT patterns for various thickness which are extracted from Figure \ref{fig: figureS8}\ (e--g).
}
\label{fig: figureS9}
\end{figure*}
